\documentclass[showkeys, groupaddress, superscriptaddress,twocolumn]{revtex4-2}
\usepackage[T1]{fontenc}
\usepackage[utf8]{inputenc}%
\usepackage{amsmath}
\usepackage{amssymb}
\usepackage{amsthm}
\usepackage{textcomp}
\usepackage{graphicx}
\usepackage[colorlinks=true, citecolor=blue, urlcolor=blue, linkcolor=black]{hyperref}
\usepackage{textgreek}
\usepackage{verbatim}
\usepackage{placeins}
\usepackage{tabularx}

\begin{document}


\title{Robust and programmable logic-in-memory devices exploiting skyrmion confinement and channeling using local energy barriers}

\author{Naveen Sisodia}
 \affiliation{Univ. Grenoble Alpes, CNRS, CEA, SPINTEC, F-38000 Grenoble, France}
 \author{Johan Pelloux-Prayer}
 \affiliation{Univ. Grenoble Alpes, CNRS, CEA, SPINTEC, F-38000 Grenoble, France}
 \author{Liliana D. Buda-Prejbeanu}
  \affiliation{Univ. Grenoble Alpes, CNRS, CEA, SPINTEC, F-38000 Grenoble, France}
 \author{Lorena Anghel}
 \affiliation{Univ. Grenoble Alpes, CNRS, CEA, SPINTEC, F-38000 Grenoble, France}
 \author{Gilles Gaudin}
 \affiliation{Univ. Grenoble Alpes, CNRS, CEA, SPINTEC, F-38000 Grenoble, France}
 \author{Olivier Boulle}
 \affiliation{Univ. Grenoble Alpes, CNRS, CEA, SPINTEC, F-38000 Grenoble, France}
 \email{olivier.boulle@cea.fr}

\ 
\date{\today}

\begin{abstract}

  Magnetic skyrmions are promising candidates for logic-in-memory applications, intrinsically merging high density non-volatile data storage with computing capabilities,  owing to their nanoscale size, fast motion,   and mutual repulsions.   However, concepts proposed so far suffer from    reliability issues as well as    inefficient   conversion of magnetic information to electrical signals. 
In this paper, we propose  a logic-in-memory device   which exploits skyrmion confinement and channeling using anisotropy energy barriers to achieve reliable data storage and synchronous shift in racetracks combined with cascadable and reprogrammable logics  relying purely on magnetic interactions.  The device combines a racetrack shift register based on skyrmions confined in nanodots with  Full Adder (FA) gates. The designed FA is reprogrammable and cascadable and can also be used to perform simple logic operations such as AND, OR, NOT, NAND, XOR and NXOR.  The monolithic design of the logic gate and the absence of any complex electrical contacts makes the device ideal for integration with conventional CMOS circuitry.

\end{abstract}

\maketitle


\section*{\label{sec:intro}Introduction}
 
“Logic-In-Memory” (LIM) architectures have  recently emerged as an  alternative to Von Neumann architectures, where the constant shuttle of data between logic and memory units leads to a critical rise in energy consumption and delay when downscaling. This has led to the search for novel technologies that combine memory and logic functionalities. Recently, magnetic skyrmions have been proposed  as the building block of such logic-in-memory technologies. Magnetic skyrmions are local chiral whirling of the magnetization. Their small lateral dimensions, down to the nanometer size and topological stability grant them particle-like properties, which, combined with the possibility to  be manipulated via electrical currents, can be exploited to code data and achieve computation at the nanoscale. These textures appear promising for logic-in-memory applications since they intrinsically merge high density non-volatile storage  and logic capabilities. However, while a number of memory and logic concepts have been proposed based on skyrmions~\cite{song2021logic,gnoli2021skyrmion,zhang2020stochastic,chauwin2019skyrmion,zhang2019skyrmion,he2017current,xing2016skyrmion,zhang2015magnetic,luo2018reconfigurable,yan2021logic16,mankalale2019skylogic,Fattouhi2021PhysRevApplied},  several major issues have hindered their technological advancement. In the first concepts of the skyrmion racetrack~\cite{fert_skyrmions_2013}, data was encoded through the  distance between neighboring skyrmions and the memory operation relies on the  synchronous shift of skyrmion trains, where the skyrmion interdistance remains constant. However, the latter can be easily perturbed by  thermal activation or   small variation in the skyrmion velocities, induced for instance by  local changes in the magnetic properties in the material or process variations.
  Different pathways have been proposed to solve this issue such as double lanes~\cite{muller_magnetic_2017} or local gate control~\cite{kang2016complementary}. However, these solutions appear difficult to implement.  For instance, local gate control requires numerous gate contacts leading to increased periphery area, limited density and increasing complexity. Regarding skyrmion based logic,  the fundamental building blocks need to fulfill a set of criteria, including cascading, fan-out, logic level restoration, immunity to noise, mitigation of potential loss of information, and input-to-output isolation. However, concepts proposed so far required complex operations and faced serious limitations. These include skyrmion-charge inter-conversion which limits the advantage of a full skyrmionic signal~\cite{mankalale2019skylogic,xing2016skyrmion,he2017current,yan2021logic16}, large  skyrmion circuits requiring complex synchronization of data~\cite{chauwin2019skyrmion}, multiple gate voltages with complex implementation~\cite{zhang2020stochastic,zhang2019skyrmion}, etc. In addition, several proposals do not fulfill the aforementioned basic requirements of logic, namely cascading, input and clock synchronization, etc \dots Thus,   there is currently no concept for a full skyrmionic computer,  incorporating logic,  memory and interconnects using exclusively magnetic signals without the need for intermediate conversion to charge signals.

 \begin{figure*}[!htb]
\includegraphics[width=0.9\linewidth]{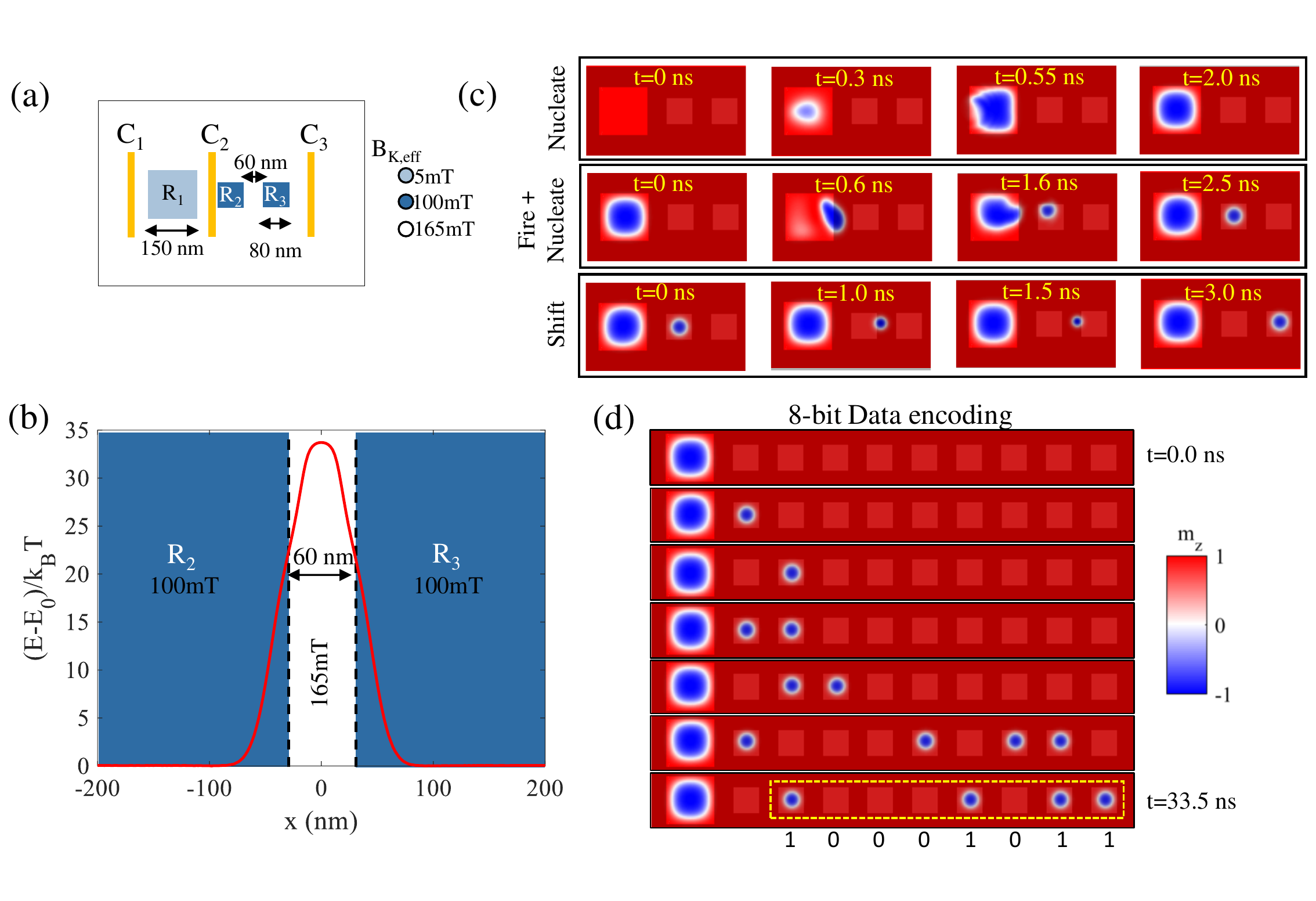}
\caption{\label{fig:nucleate}(a) Schematic of a skyrmion generator with regions of different anisotropy values. Current can be sent in-between $C_1-C_2$ or $C_2-C_3$. (b) Energy  as the skyrmion moves between two bit cells $R_2$ and $R_3$. $E_{\rm 0}$ is the energy of the skyrmions inside the bit cell. The anisotropy fields are indicated in the different regions. (c) Protocol for data encoding: a skyrmion is first nucleated in the low anisotropy region (row 1) by sending current between $C_1-C_2$ (J=15 MA/m$^2$ for 0.3 ns) and then pushed ("fired") to the 80nm cells (row 2) which work as memory bits ($J=35\rm ~MA/cm^{2}$ for $0.55\rm~ns$). A shift operation is carried out by sending current between $C_2-C_3$ (row 3) ($J=15\rm~MA/cm^{2}$ for $0.9\rm~ns$). We show only the top layer magnetization of the SAF as the bottom layer is magnetized symmetrically opposite to the top layer due to strong RKKY coupling. (d) Full 8-bit skyrmion data encoding of an 8-bit binary number ``10001011" using operations given in (c). Note that the data is encoded starting first from the least significant digit.}
\end{figure*}

Besides the aforementioned limitations, moving skyrmions are subject to the skyrmion Hall effect (skHE), namely a motion transverse to the driving force. To address this problem, topological spin textures with vanishing topological charges and thus skHE have been  proposed as  alternatives of skyrmions~\cite{zhang2016antiferromagnetic,jin2016dynamics,2018_SciRep_KolesnikovSamardakOgnev,sisodia_droplet_PRB_2021,dohi2019formation,legrand2020room,juge_skyrmions_2021,saf_zhang2016magnetic}. Of particular interest are skyrmions  in synthetic antiferromagnets (SAF), which have been recently demonstrated to be stable at room temperature~\cite{dohi2019formation,legrand2020room,juge_skyrmions_2021}. These skyrmions are  resilient against  external stray magnetic fields and possess velocities which are much higher than their ferromagnetic counterparts, making them good candidates for elementary building blocks in storage and logic. In addition to using magnetic configurations with zero topological charge, skHE can also be suppressed with different confinement techniques~\cite{indent3d_pathak,albisetti2016NatNano_nanopatterning,albisetti2017AIPnanopatterning,guang2020creating,juge2021helium,domainwallconfinement}. In particular, we have shown that skyrmion channels can be  defined by a local modification of the magnetic properties using light ion irradiation. This can be exploited to guide the skyrmion dynamics and suppress   the skyrmion Hall effect~\cite{juge2021helium}. This technique can also be leveraged for a reliable control of the skyrmion position in racetracks as well as guide their motion in complex geometries to achieve logic operations.

In this work, using micromagnetic simulations, we propose  a Logic-In-Memory (LIM) device based on SAF skyrmions which leverages skyrmion confinement using local variation of the anisotropy. It combines two main innovations: Firstly, a novel concept of  racetrack shift register where the skyrmion position is defined  by low anisotropy dots and moved reliably using current induced spin-orbit torques. This provides an elegant solution to the longstanding issue of reliability of data retention and shift operation in racetracks. We develop elementary protocols for basic ``nucleate" and ``shift" operations on a train of skyrmions. Secondly,  a compact Full Adder logic gate extendable to n-bit FA by cascading is designed. The designed FA   can be re-programmed to perform  different logic operations (e.g. NOT, BUFFER, AND, OR, XOR, NXOR, NAND). Our proposal allows  a simple and intuitive synchronization scheme   which in turn enables us to seamlessly cascade  logic design to implement large-scale networks without any additional electronic circuitry.  The designed logic architecture can also tolerate deviations in the amplitude or width of the current pulses which may arise from the electrical part of the design.

\section*{\label{sec:nucleation}Skyrmion Nucleation and Confinement}

As shown in our previous work~\cite{juge2021helium}, it is possible to artificially create an energy barrier for a skyrmion using focused  He$^+$ ion-irradiation which locally modifies the material properties (anisotropy and Dzyaloshinskii-Moriya interaction (DMI)) of the ferromagnet. Using this feature, we can engineer regions of different anisotropy values specifically tuned for two different tasks, namely  nucleation and  confinement of skyrmions.

For the micromagnetic simulations, a SAF structure is assumed composed of two  Co  layers with a thickness of 0.9nm antiferromagnetically coupled by RKKY interaction and  separated by a thin spacer [see Appendix.~\ref{sec:methods} for details regarding the micromagnetic model and parameters]. The magnetic parameters for the irradiated and non-irradiated areas are close to the ones of the Pt/Co/MgO stacks measured experimentally in Ref.~\cite{juge2021helium}. The He$^+$ ion irradiation leads to a decrease of the perpendicular magnetic anisotropy as well as the DMI.

In Fig.~\ref{fig:nucleate}(a), we show the schematic of a skyrmion generator device coupled with a memory (bit) cell which can physically confine and hold a skyrmion inside of it. There are three different regions of anisotropy. The larger $\rm150 nm$ square-shaped cell ($R_1$) with a very small effective anisotropy of $B_{\rm K,eff}=5\rm mT$ is used to nucleate the skyrmion. The anisotropy of this region is intentionally kept smaller to reduce the current density required to nucleate the skyrmion. The second region is composed of the $\rm80 nm$ bit cells ($R_2$ and $R_3$) with $B_{\rm K,eff}=100\rm mT$, which are surrounded by  high anisotropy regions of $B_{\rm K,eff}=165\rm mT$. These act as a barrier for the skyrmion, effectively trapping the skyrmion inside the bit cells. In Fig.~\ref{fig:nucleate}(b), we show the energy of the skyrmion as a function of its position $x$ as it passes from region $R_2$ to $R_3$. The energy is highest in the middle of regions $R_2$ and $R_3$. The energy barrier is $\sim33k_BT$ providing reasonable stability to the device against thermal noise. More information on the calculation and optimization of the energy barrier is given in Appendix.~\ref{sec:barrier}.

To nucleate and shift the skyrmions,  three separate contacts $C_1$, $C_2$ and $C_3$  are used  (Fig.~\ref{fig:nucleate}(a)). 
To nucleate a skyrmion in $R_1$ (first panel of Fig.~\ref{fig:nucleate}(c)),   a short ($0.3\rm~ns$) current pulse  ($J=15\rm ~MA/cm^{2}$) is first injected in  between contacts $C_1-C_2$. A second current pulse with a higher current density  ($J=35\rm ~MA/cm^{2}$ for $0.55\rm~ns$)   allows the skyrmion to overcome the barrier and move from $R_1$ towards $R_2$. Simultaneously, due to the high current density, another skyrmion nucleates in the region $R_1$. This whole operation can thus be dubbed as ``fire+nucleate" as one skyrmion is nucleated and another skyrmion is simultaneously fired (panel 2 of Fig.~\ref{fig:nucleate}(c)). Once nucleated, the  shift operation is performed  by injecting a current between contacts $C_2$ and $C_3$ such that the skyrmions move between $R_2-R_3$ without influencing the nucleation region $R_1$ (third row panel in Fig.~\ref{fig:nucleate}(c)). Overall, to encode a full data stream, the following protocol is adopted: 1.) Nucleate in region $R_1$, 2.) If Bit 1 is required, use ``fire+nucleate" followed by ``shift" 3.) If Bit 0 is required, use only ``shift". We show in Fig.~\ref{fig:nucleate}(d) and in video  SV1 the encoding of an 8-bit binary number ``10001011" in a racetrack composed of 9 cells using this protocol. After nucleation, the skyrmion train shifts synchronously along the cell track when injecting the current pulses. Thus, our proposed racetrack design based on confinement cells and current induced tunneling provides large bit stability and  natural synchronization of the bit train, solving an important issue of initial skyrmion racetrack design.    

\begin{figure*}[!htb]
\includegraphics[width=0.7\linewidth]{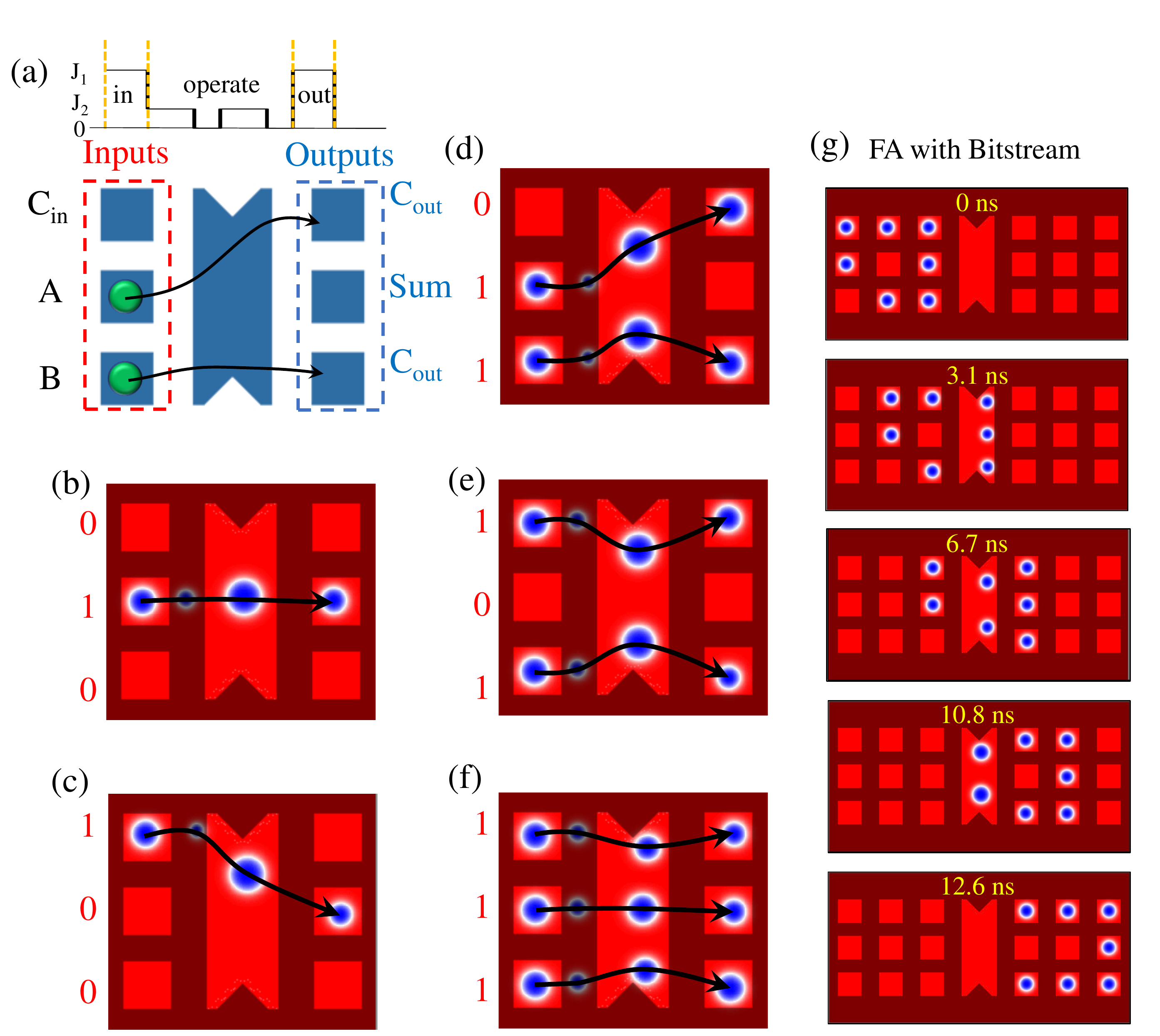}
\caption{\label{fig:1bit}Design and operation of a 1-bit Full-Adder (FA) based on irradiated SAF. (a) shows the schematic of the FA and the current pulses which are designed to move the skyrmions from their respective bits on the left to the middle region where the operation takes place and then to the output bits on the right. The current pulses $J_1$ and $J_2$ have an amplitude and duration  of   $J_1=15\rm~MA/cm^{2}$ and $t_1=0.8\rm~ns$ and $J_2=5\rm~MA/cm^{2}$ and $t_2=0.9\rm~ns$.  (b)-(f) show the micromagnetic simulation of the 1-bit FA gate for various test cases. Snapshot of the magnetization at different times are overlayed to show the complete trajectories of the skyrmions (black solid curves). (g) Serial FA operation on a stream of $C_{in}$, $A$ and $B$ inputs. }
\end{figure*}

\section*{\label{sec:1BitFA}Full Adder Design}

 A Compact full Adder (FA) logic gate can also be designed using anisotropy energy barriers in irradiated SAF stacks. A Full Adder is considered as the basic unit in Arithmetic Logic Unit (ALUs) performing bitwise addition of two binary numbers. It has three inputs, $A$, $B$ and $carry-in~(C_{\rm in})$ and produces two outputs, $Sum$ and $carry-out~(C_{\rm out})$.  The logic gate is divided into three parts (see Figure.~\ref{fig:1bit}(a)).  The left part of the device is made of three bit cells, which are the inputs $A$, $B$ and $C_{\rm in}$. Similarly, at the right-hand side of the logic gate, three other bit cells  hold the two outputs $Sum$ and $C_{\rm out}$ of the FA. The middle region of the design represents the part of the device which performs the logic operation. All three regions and the individual bit cells are separated from each other by anisotropy energy barriers.

We show the operation of the designed Full Adder gate  in Figs.~\ref{fig:1bit}(b)-(f) and in video SV2:  The skyrmions which are present inside the bit cells inputs $A$, $B$ and $C_{\rm in}$ on the left region are pushed inside the middle region after passing over the energy barrier by sending a  current pulse  uniformly through the  device ($J_1=15\rm~MA/cm^{2}$). When skyrmions are inside the middle region, different possibilities arise depending on the total number of skyrmions inside the middle region. For only one skyrmion [Figs.~\ref{fig:1bit}(b)-(c)], the most stable state for the skyrmion is to stay at the center of the middle region. However, when there are two skyrmions [Figs.~\ref{fig:1bit}(d)-(e)], they repel each other and would stay near the top and bottom edge of the middle region to minimize their mutual interaction. When the total number of skyrmions in the middle region is three [Figs.~\ref{fig:1bit}(f)], the skyrmions will maximize the individual distances between them and will acquire a position at the top, middle and bottom parts. 
To achieve the relaxation of the interacting skyrmions in the middle part, two pulses with smaller amplitude $J_2=5\rm~MA.cm^{-2}$  are then injected. The skyrmions are then pushed into the rightmost region by sending a current pulse with a larger amplitude $J_1$. At the output end, the middle cell will only have a skyrmion if the total number of skyrmions on the input side is either one or three. This represents the $Sum$ output of the Full Adder. Similarly, the top and bottom outputs will only have a skyrmion if the total number of input skyrmions is either two or three. This is the same as the $C_{\rm out}$ (carry-out) output of the Full Adder. We thus achieve an entire full adder operation with a single logic gate.

We have specifically designed our FA logic gate in Fig.~\ref{fig:1bit}(a) keeping in mind an easy integration with racetrack storage  such as the one in Fig.~\ref{fig:nucleate}(a). Using this architecture, the FA logic gate can be easily extended to perform logic operations with a continuous stream of data to serially compute $Sum$ and $C_{\rm out}$ outputs for each input set of data bits. Such an operation is shown in Fig.~\ref{fig:1bit}(g), where we show serial 1-bit FA calculation on streams of input $A$, $B$ and $C_{\rm in}$ [see also video SV3]. The current pulses, in general, follow the same protocol and can be easily extended to perform $n$ operations, where $n$ is the total number of bits in the bit-stream.\footnote{The only modification needed in the current pulses is the merging of $J_1$ pulse of outgoing operation with the $J_1$ pulse of incoming operation as both of these operations can be carried out simultaneously without affecting the logic operation.}

\begin{figure*}
 \includegraphics[width=\linewidth]{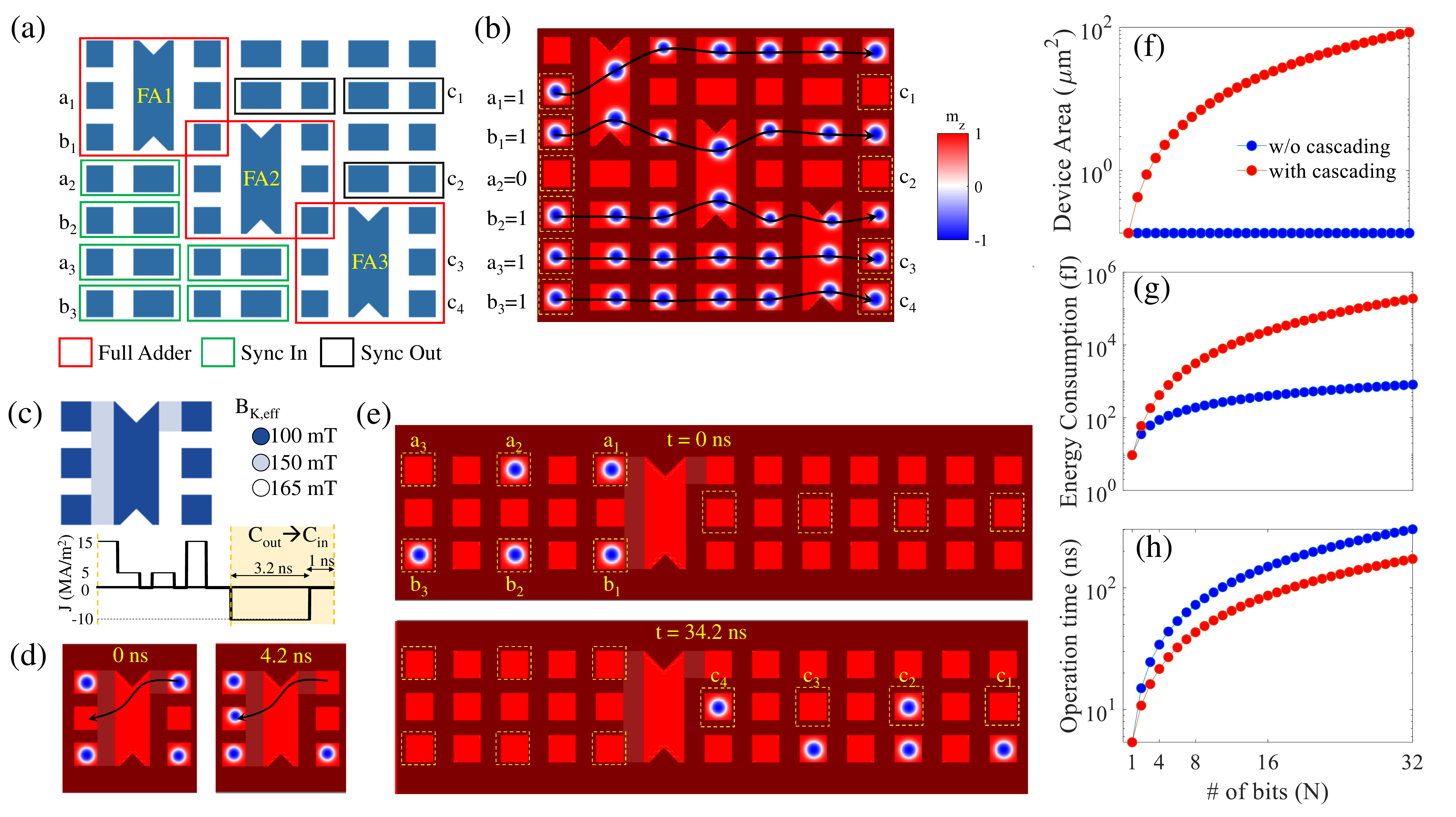}
\caption{\label{fig:3bit}(a) Design of a 3-bit Full-Adder (FA) by cascading three 1-bit FAs. Additional components have been added to synchronize inputs application and the entire operation (b) shows the micromagnetic simulation for the case of (101+111=1100). The current pulses used to perform the logic operation are exactly the same as the 1-bit FA operation repeated 3-times. (c) and (d) show the design and operation of an FA where the carry-bit can be sent back using current of negative polarity with density $J=-10\rm~MA/cm^{2}$ for $t=3.2\rm~ns$ followed by an off-pulse ($J=0\rm~MA/cm^{2}$) of $1\rm ~ns$. (e) Operation of a 3-bits FA using the modified FA shown in (c). The operation corresponds to the case (011+111=1010). (f)-(h) shows the comparison of both approaches for 3-bits addition in terms of device area, energy consumption and operation time, respectively.
}
\end{figure*}

\section*{\label{sec:3bitFA}Full Adder Multi-bit operation}

One limitation of the FA gate is that it only operates on 1-bit inputs. For instance,  the actual operation on the bitstreams shown  in Fig.~\ref{fig:1bit}(g)  is  a 1-bit FA operation done serially on each arriving set of input bits, since the carry-bit ($C_{\rm out}$) is not taken forward in each successive computation. We show here that $N$-bit FA gates can be easily designed by properly cascading 1-bit FA gates. As an example, we present in Fig.~\ref{fig:3bit}(a) the design of a 3-bit FA gate composed of  three cascaded FA gates where the carry bit from each computation is used in the next computation. The bottom $C_{\rm out}$ from the FA1 is connected to the top bit cell of FA2, such that the output $C_{\rm out}$ from FA1 operation is used as one of the inputs of FA2 operation. To ensure that the two other  inputs of FA2 reach  at the same time as the output $C_{\rm out}$ of FA1, we add some delay gates Sync-In (green boxes) before FA2. These synchronization gates can  be cascaded  to vary the delay time and provide on-demand synchronization between various segments of logic operations. Similar gates are also added to synchronize the outputs (Sync-Out, black).

In Fig.~\ref{fig:3bit}(b) and video SV4, we show the operation of a 3-bit FA using cascaded 1-bit FAs. The calculation performed is ($a_3~a_2~a_1$) +  ($b_3~b_2~b_1$) = ($c_4~c_3~c_2~c_1$), where ($a_3~a_2~a_1$)=(101) and ($b_3~b_2~b_1$)=(111). The expected output is ($c_4~c_3~c_2~c_1$)=(1100) which is correctly obtained by bit cells $c_4$, $c_3$, $c_2$ and $c_1$ shown by dotted yellow boxes.
 
 The design of Fig.~\ref{fig:3bit}(a) is very promising for fast n-bit FA operation and for further modifications using cascading and synchronization. However, a more compact and low power design can be  implemented by  directly computing the n-bits FA operation using  streams of skyrmions in two racetracks and a single FA gate. This would be similar to the operation of FA shown in Fig.~\ref{fig:1bit}(g), however, with bit-streams replaced with $N$-bit numbers. For this purpose, we modify our FA design to include another region of different anisotropy value as shown in Fig.~\ref{fig:3bit}(c). This new region (light blue) which covers the area joining the input bits to middle region and the area to the left of top $C_{\rm out}$ bit has an anisotropy in-between the anisotropy of the regular bit cells and the barrier region. This enables us to send a negative polarity current pulse which is enough to move only the $C_{\rm out}$ bit while keeping the other skyrmions at their respective places. This pulse is shown alongside in Fig.~\ref{fig:3bit}(c) [highlighted part]. In Fig.~\ref{fig:3bit}(d), we show how the $C_{\rm out}$ bit is moved back to the input end by this current pulse of negative polarity. During the entire operation, none of the other skyrmions are affected. Using this, we can now compute the addition operation of two $N$-bit numbers by successively sending them through the top and bottom inputs of the FA (middle input is left empty). The design and operation for a 3-bit FA using this strategy is shown in Fig.~\ref{fig:3bit}(e) and video SV5. It may be noted that  we need to add an empty bit between each successive bit as no operation can take place during the movement of $C_{\rm out}$ bit using negative polarity current. The snapshots in Fig.~\ref{fig:3bit}(e) correspond to the initial and final stage of the addition operation ($a_3~a_2~a_1$) +  ($b_3~b_2~b_1$) = ($c_4~c_3~c_2~c_1$), where, ($a_3~a_2~a_1$)=(011), ($b_3~b_2~b_1$)=(111) and the expected output is ($c_4~c_3~c_2~c_1$)=(1010) which is correctly computed with our implementation. The inputs and outputs are shown in dotted yellow boxes. 

 \begin{figure*}[!htb]
\includegraphics[width=0.7\linewidth]{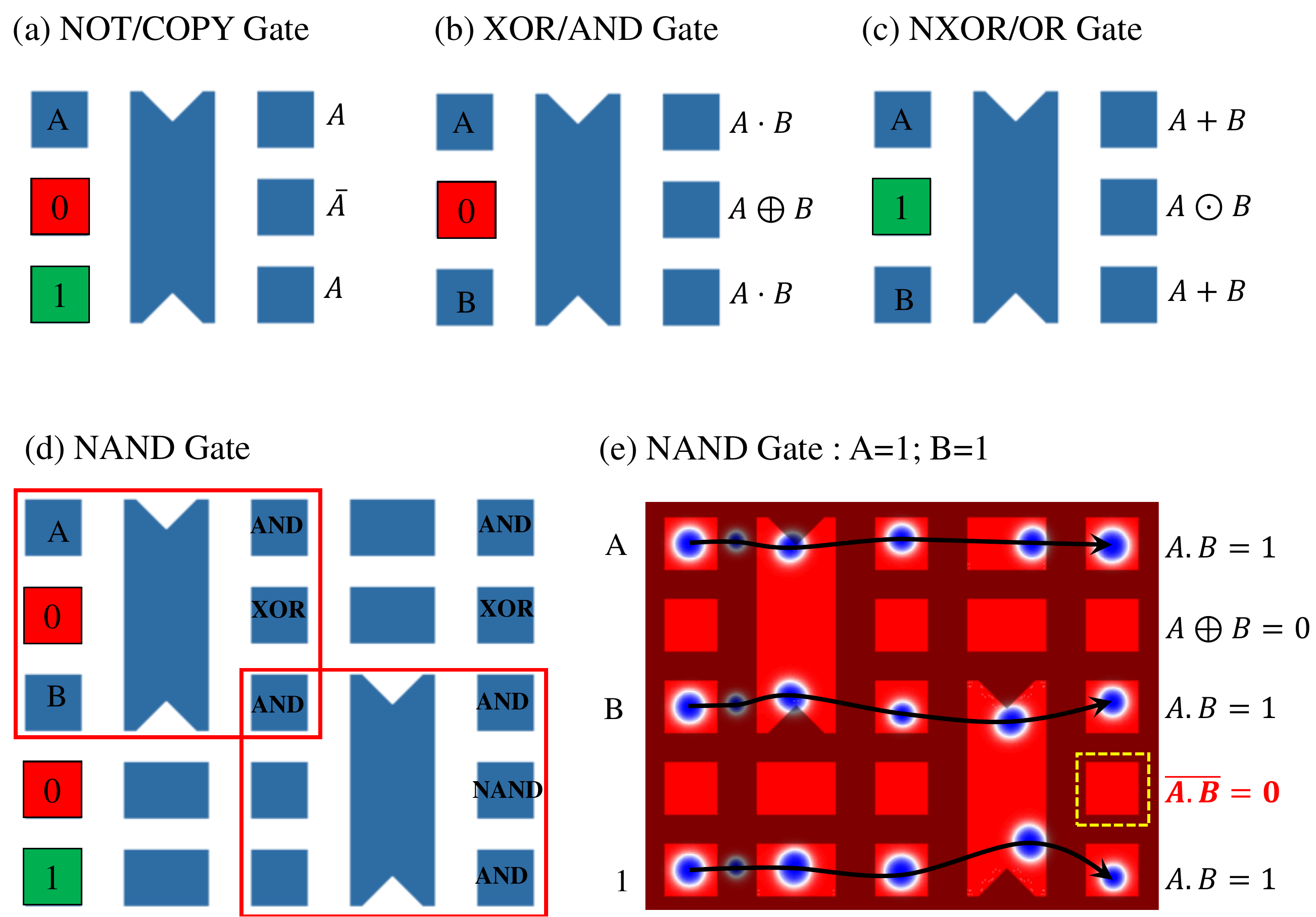}
\caption{\label{fig:nand}Different logic gate designs derived from the FA. (a) shows a NOT/COPY gate design. Some input bits have been fixed to either "0" or "1". Similarly, (b) and (c) show XOR/AND and NXOR/OR gates derived from FA. (d) shows a universal NAND gate designed by cascading AND and NOT gates, both of which are modified versions of FA. (e) shows the operation of NAND for both inputs as 1.
}
\end{figure*}

The performance of the two methods for $N$-bit FA computation described above can be compared  on the basis of three important characteristic metrics: (i) Device Area, (ii) Energy consumption and (iii) Operation time. In Fig.~\ref{fig:3bit}(f), we show the variation of the device area with the number of bits ($N$). For the cascaded FA shown in Fig.~\ref{fig:3bit}(a), both dimensions in the $x-y$ plane increase linearly with $N$ leading to an `$\mathcal{O}\rm~(N^2)$ variation of device area. However, since only one FA gate is needed for the alternative approach (without cascading) shown in Figs.~\ref{fig:3bit}(c)-(e), the device area remains the same as `$N$' increases. Note that in Fig.~\ref{fig:3bit}(e), the increase in the length along x-axis is not included in the calculation as the extended regions are considered as a part of the racetrack storage rather than logic. In Fig.~\ref{fig:3bit}(g), we show the total energy consumption of the FA logic gate as a function of number of bits. For a 1-bit operation, the energy of the operation is $9\rm~fJ$. In the case of cascaded FA operation, the energy dramatically increases with the number of bits as $\mathcal{O}\rm~(N^3)$ due to linear increase in operation time and quadratic increase in the device area. However, when using the FA design shown in Fig.~\ref{fig:3bit}(c)-(e), the total energy only increases linearly with the number of bits. For a 32-bit operation, the total energy consumption is $\sim0.8~\rm pJ$ which is more than two orders lower than the cascaded FA operation. However, this design suffers from a larger operation time compared to cascaded FA gate since each operation is delayed by $4.2\rm~ns$ to move the carry-out bit back to the input position [Fig.~\ref{fig:3bit}(h)]. Due to the Joule dissipation in the metallic conductors, the proposed device consumes 1-2 orders of magnitude more energy for individual logic operations compared to state-of-art CMOS full adder~\cite{energycomp}. However, owing to  its intrinsic logic-in-memory approach, a large gain in energy and delay is expected at the device level, since it minimizes the energy dissipated in interconnects  and the stand-by power. Note that the speed of both designs can be further increased by using skyrmions with higher velocity ($>1000\rm m/s$ is expected for SAF~\cite{tomasello2017performance} which is 6 times the maximum velocity of skyrmions in this work).

\section*{\label{sec:derivative}Re-programming FA logic}
Full adder logic gates can be easily  modified  in order to build a number of basic logic functions, which in turn can be combined to perform complex logic operations.
The output of the full adder gate is given as: \(Sum=C_{\rm in}\oplus (A \oplus B)\) ; \(C_{\rm out}= AB+BC_{\rm in}+AC_{\rm in}\). By fixing one or more of the inputs, different logic gates can be achieved. Note that the inputs of a FA logic gate are all interchangeable, so it does not matter which of the inputs are fixed. Figures~\ref{fig:nand}(a)-(c) show the modified FA gates which can perform COPY, NOT, AND, XOR, OR and XNOR operations.

 The universal NAND gate can also be designed by cascading two modified FA gates [Fig.~\ref{fig:nand}(d)]. The first gate is modified by setting one of the inputs to ``0" such that the $C_{\rm out}$ represents the output of an AND gate. The second FA gate is modified to perform an inversion operation by setting two of the inputs to ``0'' and ``1''. The operation of this gate for a test case of $A=1$ and $B=1$ is shown in Fig.~\ref{fig:nand}(e). The output $\overline{A.B}=0$ corresponding to the NAND gate is highlighted. It may be noted that we also obtain AND and XOR outputs along with the NAND output. 

\section*{\label{sec:FAtolerance}Electrical Tolerances and failure mechanisms}
\begin{figure}[!htb]
\includegraphics[width=\linewidth]{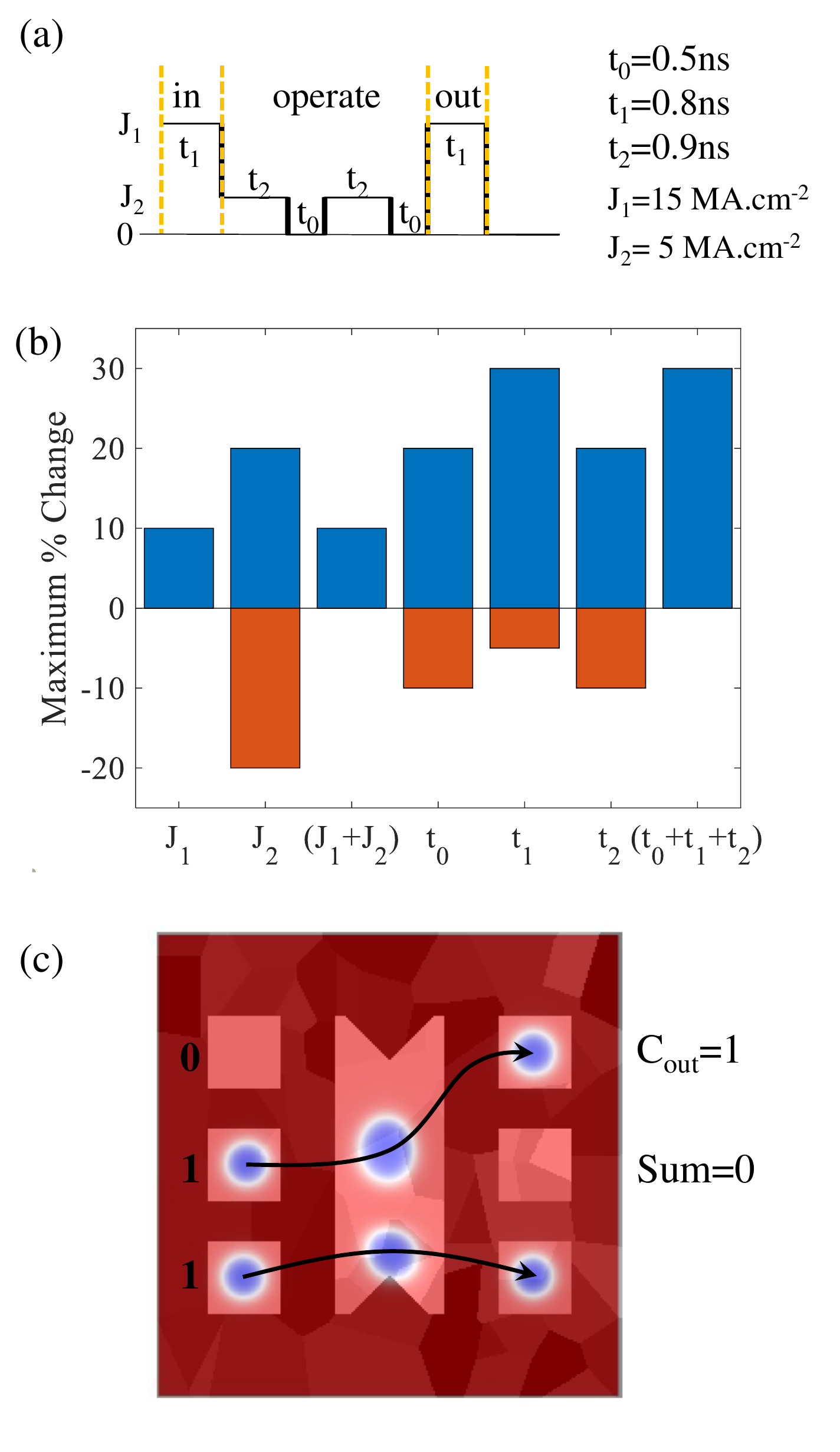}
\caption{\label{fig:tolerance}(a) Current pulses used for FA operation in Fig.~\ref{fig:1bit}. (b) Acceptable change (\%) in different parameter values in the current pulse. (c) Test case with inputs (0,1,1) for FA operation with random anisotropy grains.}
\end{figure}

For practical application of our design, it is important to probe its tolerance against variations in the current amplitude or pulse width. In Fig.~\ref{fig:tolerance}(a), we show the set of pulses used for the 1-bit Full Adder operation. There are two different current density values used in the circuit : $J_1=15\rm~MA/ cm^{2}$ and $J_2=5\rm~MA /cm^{2}$ with pulse widths of  $t_1=0.8\rm~ns$ and $t_2=0.9\rm~ns$, respectively. There is also an off-pulse of width $t_0=0.5\rm~ns$. We individually vary each of these parameters by fixing other parameters at their reference values. We also try to vary $J_1$, $J_2$ or $t_0$, $t_1$ and $t_2$, simultaneously. The results are shown in Fig.~\ref{fig:tolerance}(b) depicting the acceptable level of variation for each case for which the logic operations are executed without any errors. Note that these results are also valid for the NAND as well as other derived logic gates. These tolerance values provide valuable input for further modeling with large-scale electrical circuit designing tools.

We also performed simulations to check the robustness of device against variations in anisotropy. We varied the effective anisotropy field of the barrier region and found that the Full Adder works correctly for the range of values $\rm 155mT-167mT$. Further, we also perform simulations by adding grains of size $\rm 15nm$ with random $0.5\%$ variation of anisotropy constant (equivalent to $\sim5\%-8\%$ change in effective anisotropy field), similar to  previous works~\cite{juge2018magnetic,gross_skyrmion_2018} and show that the Full Adder works correctly. The motion of skyrmions for a test case with inputs (0,1,1) is shown in Fig.~\ref{fig:tolerance}(c). However, we note that a stronger anisotropy variation may lead to incorrect operations. In our device, this happens as the skyrmions are stabilized for the parameter values which are close to in-plane to out-of-plane transition regime. Therefore, a small change in the anisotropy constant rapidly changes the skyrmion energy and hence its trajectory. A possible strategy to mitigate this is to use higher anisotropy thin films as the pristine sample and then irradiating them to obtain the tracks/barriers with the same anisotropy difference as used in this work.

An important part of the logic gate design involves device testing against a possible failure during operation. To investigate this point, we include thermal noise corresponding to $\rm T=350~K$ in our simulations \footnote{As we are already doing simulations with experimental parameters extracted at T=300K, we only include thermal noise corresponding to T=50K to avoid overestimating the effect of temperature}. We simulate 20 instances for each of the possible 8 test cases for the Full Adder (total 160 attempts). Despite high thermal noise, the Full Adder worked correctly for $\sim85\%$ of the total attempts. For the cases in which it failed, three different failure mechanisms were found. The first one is caused by the annihilation of  input skyrmions during the operation. We predict that a stronger RKKY coupling could counter the thermal noise and improve the overall stability of the SAF skyrmions. The second failure mechanism is due to  the presence of skyrmions at the wrong output after the operation has ended. This happens when the skyrmion diffusion due to thermal noise significantly perturbs the intended trajectory of the skyrmions. Higher Gilbert damping is expected to minimize this effect. However, the trade-off will be an increased energy consumption as higher current densities will be required to move the skyrmions. 
The third possibility is a skyrmion getting stuck in between two bit cells. This occurs when the skyrmion finds a local minima in between two memory bits. Our barrier has already been optimized to avoid this situation [see Appendix.~\ref{sec:barrier}], however, further optimization is needed to comply with high-temperature conditions. This will involve increasing the difference of anisotropy between the low and high anisotropy regions which will increase the barrier height.

\section*{\label{sec:future}Conclusion}
To conclude, we proposed a logic-in-memory device based on skyrmion confinement and channeling by anisotropy energy barriers in synthetic antiferromagnets. We first designed a racetrack  shift  register memory   based on skyrmions  confined by anisotropy energy barriers as memory bits, allowing reliable data storage with  large stability and robust synchronous shift operations on skyrmion train. This design naturally solves  the stability and synchronization issues of the initial racetrack concept without introducing additional gates or complex geometries.   We then combine our racetrack storage with a newly designed 1-bit Full Adder (FA) gate extendable to $N$-bits FA by cascading. The designed FA is reprogrammable and can also be used to perform AND/OR/NOT/NAND/XOR/NXOR operations. The device is expected to perform well even with some fluctuations in the amplitude/width of the injected current pulses and in presence of thermal noise. The simplicity and compactness of the design combined with the minimal use of electrical circuitry brings the proposed device to the same level as the current technological capabilities of fabrication/manufacturing processes, thus providing an important advance for the  development of skyrmion-based logic-in-memory technology.

\begin{acknowledgments}
The authors acknowledge financial support from the French national research agency (ANR) (Grant Nos. ANR-15-CE24-0015-01 and ANR-17-CE24-0045) and the American defense advanced research project agency (DARPA) TEE program (Grant No. MIPR HR0011831554). This work has been partially supported by MIAI@Grenoble Alpes, (ANR-19-P3IA-0003).
\end{acknowledgments}

\appendix

\section{\label{sec:methods}Micromagnetic Simulation Model}
We use the micromagnetic package $Mumax^3$~\cite{VansteenkisteWaeyenberge_AIPadv2014} to simulate the magnetization dynamics of a Synthetic Antiferromagnet having a Heavy Metal (HM)/Ferromagnet (FM)/Spacer (S)/Ferromagnet (FM) structure by solving Landau Lifshitz Gilbert (LLG) equation with an additional Spin-Orbit Torque (SOT) term.
\newcommand{\Mi}{\textbf{m}_{\rm i}}

\begin{gather}\label{eq:llgs}
\begin{split}
\frac{d\Mi}{dt}=&-\gamma (\Mi\times \textbf{B}_{\rm eff, i}) + \alpha(\Mi\times \frac{d\Mi}{dt})\\
&-\mathcal{\eta} \Big(\Mi\times(\Mi\times\textbf{e}_{\rm p})+\epsilon^{'}(\Mi\times\textbf{e}_{\rm p})\Big)\\
\mathcal{\eta}=&\frac{\hbar\theta_{\rm H} J }{2|e|M_{\rm S}d}
\end{split}
\end{gather}

Here $\Mi$ is the normalized magnetic moment of $i^{\rm th}$ ferromagnetic layer. $B_{\rm eff,i}$ is the effective field on each of the layers which contain contributions from the anisotropy, exchange, DMI and magnetostatic interactions along with the additional field to account for RKKY coupling between layers. We assume that SOT acts only on the top ferromagnetic layer of the SAF which is in contact with a Heavy Metal which generates a spin-orbit torque. The spin Hall angle is assumed to be $\theta_{\rm H}=0.4$. The ratio of field-like torque to damping-like torque is $\epsilon^{'}=0.01$. The RKKY coupling strength between FM layers is $-0.2\rm~mJ/m^2$. The thickness of each layer in the FM/S/FM trilayer structure is $0.9\rm~nm$ (HM not included in micromagnetic simulations). The simulation region is divided into micromagnetic cells of size $\rm 2~nm\times2nm\times0.9nm$.
Unless specified explicitly, we use the two sets of parameters corresponding to irradiated and non-irradiated regions of the SAF, given in Table~\ref{tab:parameter}. These parameters are similar to our previous experimental work on $He^+$-ion irradiated Co-based FM~\cite{juge2021helium}. However, we have assumed higher anisotropy values to stabilize the skyrmions at zero-field, while keeping the difference between the anisotropy values of irradiated and non-irradiated region same. We also note that for the He ion dosage ($<2\times10^{14}\rm ions/cm^2$) required to create the anisotropy barriers used in this work, the irradiation is expected to change the resistivity of Heavy Metal which generates the SOT (Pt in the reference experiments~\cite{juge2021helium}) by $<1\%$. Therefore, we do not expect any changes in the dynamics of the skyrmions due to non-uniform current flow.

\begin{table}[t]
\caption{\label{tab:parameter}%
Simulation Parameters
}
\begin{ruledtabular}
\begin{tabular}{lcc}
\textrm{Parameter}&
\textrm{Irradiated}&
\textrm{Non-irradiated}\\
\colrule
$M_s~\rm (MA/m)$ & 1.32 & 1.32\\
$B_{\rm K,eff}~\rm (mT)$ & 100 & 165\\
{$\vert{}D\vert~\rm (mJ/m^2)$} & 1.082 & 1.122\\
$A\rm~(pJ/m)$ & 16 & 16\\
${\alpha}$ & 0.05 & 0.05\\
\end{tabular}
\end{ruledtabular}
\end{table}

\section{Energy Barrier Calculation\label{sec:barrier}}

\begin{figure}[h]
\includegraphics[width=\linewidth]{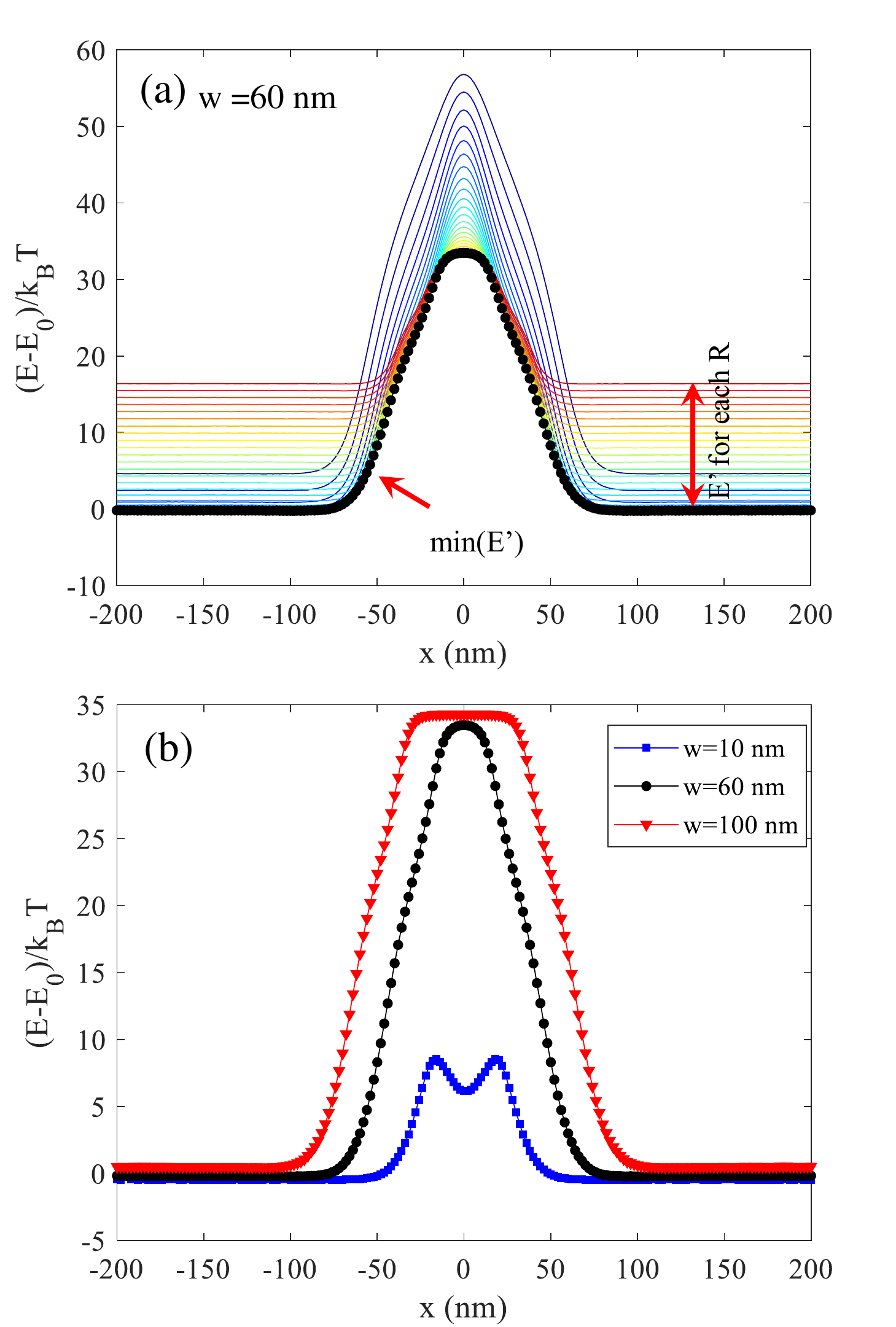}
\caption{\label{fig:barrier}(a) Energy of a rigid skyrmion passing through the barrier as a function of its position for different values of skyrmion radius. (b) Barrier energy as a function of skyrmion position for different values of barrier widths, $w$.}
\end{figure}
The energy of the skyrmion crossing the anisotropy barrier is dependent on the position of the skyrmion as well as the skyrmion radius: $ E^{'}=f(x,R)$. From our simulations, we calculate the total micromagnetic energy for all values of $x$ and $R$ for a particular barrier width $w$. We then reduce the map $\rm E^{'}$ using $E=min(E^{'})=min[(f(x,R)]$ taking the minimum along the $R$ dimension. We show $E^{'}$ scans along $x$ with varying $R$ in Fig.~\ref{fig:barrier}(a). The curve corresponding to minima of $E^{'}$ is shown by solid black circles. This corresponds to Fig. 1(b) in the main text for $w=60\rm~nm$. In Fig.~\ref{fig:barrier}(b), we show the anisotropy barrier as a function of the width of the barrier. For low barrier width $w=10\rm~nm$, the barrier has lower peak value ($<10k_BT$). Moreover, the barrier also has a local minima at the center. This is not desirable for skyrmion confinement as there is a possibility of skyrmion getting stuck in the middle of the barrier during logic operation. For high barrier width ($w=100\rm~nm$), while the barrier height is sufficient ($\sim33k_BT$), the shape of the barrier is flat around the center region. This is also an unfavourable situation as in the absence of current, there will be no driving force on the skyrmion to push it towards the regions of confinement. In comparison, $w=60\rm~nm$ (which is used in main text) has an ideal shape. Due to the sharp peak at the center, the barrier is always forcing the skyrmions to move towards lower anisotropy regions where they can be confined.

\section{4-input Logic Device}
\begin{figure}[h]
\includegraphics[width=0.6\linewidth]{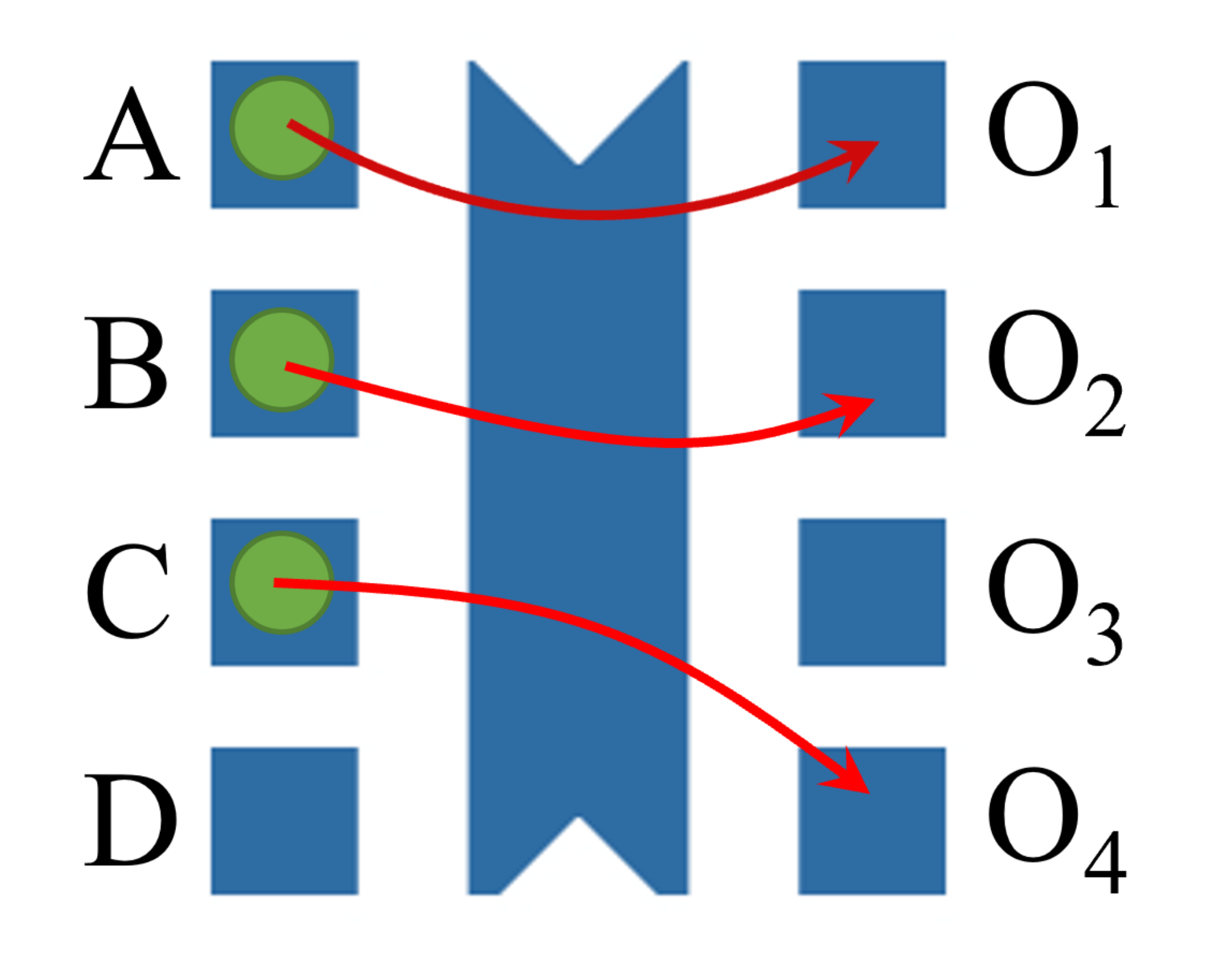}
\caption{\label{fig:4input}(a) Illustration of a 4-input logic gate}
\end{figure}

The FA design shown in the main text can also be used as a template to design logic gates with more than 3 inputs. We show one such example in Fig.~\ref{fig:4input} which supports 4 skyrmion inputs. Using the same protocol as for the FA gate, we perform full micromagnetic simulations to obtain the truth table corresponding to the logic functionality of the 4-input gate [Table ~\ref{tab:truth}]. This truth table can be used to customize the logic gate to calculate different logic functions. For example, if the input A is set to 0, the output of $O_2$ represents the operation $B\oplus(C\cdot D)$.

\begin{table}[h]
    \centering
    \begin{tabular}{cccc|cccc}

\hline

  $A$ & $B$ & $C$ & $D$ & $O_1$ & $O_2$ & $O_3$ & $O_4$\\ \hline \hline
  0 & 0 & 0 & 0 & 0 & 0 & 0 & 0 \\
  1 & 0 & 0 & 0 & 0 & 1 & 0 & 0 \\
  0 & 1 & 0 & 0 & 0 & 1 & 0 & 0 \\
  0 & 0 & 1 & 0 & 0 & 0 & 1 & 0 \\
  0 & 0 & 0 & 1 & 0 & 0 & 1 & 0 \\
  0 & 0 & 1 & 1 & 0 & 1 & 0 & 1 \\
  0 & 1 & 1 & 0 & 0 & 1 & 1 & 0 \\
  1 & 1 & 0 & 0 & 1 & 0 & 1 & 0 \\
  1 & 0 & 1 & 0 & 0 & 1 & 1 & 0 \\
  0 & 1 & 0 & 1 & 0 & 1 & 1 & 0 \\
  1 & 0 & 0 & 1 & 0 & 1 & 1 & 0 \\
  0 & 1 & 1 & 1 & 1 & 0 & 1 & 1 \\
  1 & 0 & 1 & 1 & 1 & 0 & 1 & 1 \\
  1 & 1 & 0 & 1 & 1 & 1 & 0 & 1 \\
  1 & 1 & 1 & 0 & 1 & 1 & 0 & 1 \\
  1 & 1 & 1 & 1 & 1 & 1 & 1 & 1 \\
  \hline
  
\end{tabular}
    \caption{Truth table for the 4-input logic gate}
    \label{tab:truth}
\end{table}

\section{Supplementary Videos}
\begin{itemize}
    \item SV1 : 8-bit data encoding using ``Fire+Nucleate" and ``Shift" operations
    \item SV2 : Operation of 1-bit FA for all test cases
    \item SV3 : 1-bit FA operating on a stream of inputs A, B and $C_{in}$
    \item SV4 : 3-bit addition using three cascaded FA gates
    \item SV5 : 3-bit addition using a single modified FA gate
    
\end{itemize}

\bibliography{apssamp}

\end{document}